\documentclass[useAMS,usenatbib]{mn2e}

\usepackage{graphicx}

\def\rc{\mbox{$R_{\rm c}$}}
\def\rt{\mbox{$R_{\rm t}$}}
\def\rap{\mbox{$D_{\rm app}$}}
\def\ms{\mbox{$M_\odot$}}

\title[A catalogue of extended objects in the MCs]{A general catalogue of extended objects in the 
Magellanic System}

\author[E. Bica et al.]{E. Bica$^1$, C. Bonatto$^1$, C.M. Dutra$^2$, and J.F.C. Santos Jr.$^3$\\
$^1$ Departamento de Astronomia, Universidade Federal do Rio Grande do Sul, Av. Bento Gon\c{c}alves 9500\\
Porto Alegre 91501-970, RS, Brazil\\
$^2$ Universidade Federal do Pampa - UNIPAMPA, Centro de Ci\^encias da Sa\'ude, Rua Domingos de Almeida, 3525,\\
Bairro S\~ao Miguel, Uruguaiana 97500-009, RS, Brazil\\
$^3$ Departamento de F\'\i sica, ICEx, Universidade Federal de Minas Gerais, Av. Ant\^onio Carlos 6627\\
Belo Horizonte 30123-970, MG, Brazil }

\begin{document}

\pagerange{\pageref{firstpage}--\pageref{lastpage}}

\maketitle

\label{firstpage}

\begin{abstract}
We update the SMC, Bridge, and LMC catalogues of extended objects that were constructed by
members of our group from 1995 to 2000. In addition to the rich subsequent literature for 
the previous classes, we now also include HI shells and supershells. A total of 9305 objects 
were cross-identified, while our previous catalogues amounted to 7900 entries, an increase of
$\approx12\%$. We present the results in subcatalogues containing 1445 emission nebulae, 3740 
star clusters, 3326 associations, and 794 HI shells and supershells. Angular and apparent size 
distributions of the extended objects are analysed. We conclude that the objects, in general, 
appear to respond to tidal effects arising from the LMC, SMC, and Bridge. Number-density profiles 
extracted along directions parallel and perpendicular to the LMC bar, can be described by two 
exponential-disks. A single exponential-disk fits the equivalent SMC profiles. Interestingly, 
when angular-averaged number-densities of most of the extended objects are considered, the 
profiles of both Clouds do not follow an exponential-disk. Rather, 
they are best described by a tidally-truncated, core/halo profile, despite the fact that the 
Clouds are clearly disturbed disks. On the other hand, the older star clusters taken isolately, 
distribute as an exponential disk. The present catalogue is an important tool for the unambiguous
identification of previous objects in current CCD surveys and to establish new findings.
\end{abstract}


\section{Introduction}
\label{intro}

The Magellanic Clouds are fundamental galaxies for astrophysics owing e.g. to their proximity, chemical 
compositions, age distributions, star cluster structural properties and related dynamical evolution, and 
as two close-by interacting galaxies  (e.g. \citealt{Wes90}, \citealt{DaCosta91}, \citealt{Piatti02},
\citealt{MG03}, \citealt{DMG02}, \citealt{BekChi07}). \citet{Schaefer08} reviews recent estimates of the 
distance to the Clouds and arrives at the values $d_{\rm LMC}\approx50$\,kpc and $d_{\rm SMC}\approx60$\,kpc.

More than 50 years have elapsed since the first attempts to systematically catalogue extended objects in 
the Magellanic Clouds (e.g. the nonstellar emission nebulae of \citealt{Henize56}). We refer as extended
objects the emission nebulae, star clusters, associations, and HI shells and supershells. Early catalogues 
of star clusters included brighter ones in the SMC (\citealt{Kron56}, \citealt{Lindsay58}) and LMC 
(\citealt{ShaLind63}; \citealt{LynWes63}). Fainter clusters with deeper photographic material were detected,
e.g. by \citet{HodSex66} and \citet{Hodge86}. Binary or multiple clusters are another characteristics of many
Magellanic Cloud clusters, showing their importance for cluster dynamical evolution (\citealt{BhaHad88}; 
\citealt{Oliveira00}). Examples of catalogues of associations are \citet{LuHo70} for the LMC, \citet{Hodge85}
for the SMC, and \citet{BatDem92} in the Bridge. A complementary study to \citet{Henize56} is the catalogue of
nebular complexes by \citet{DEM76} based on H$\alpha$ plates.

Hubble Space Telescope (HST) instrumention revealed serendipitously two faint clusters in the LMC bar that were 
undetected in Sky Survey plates (\citealt{Santiago98}). These two clusters suggested the existence of an important
undetected faint population of clusters. CCD mosaics, e.g. \citet{Pietrzynski98} in the SMC central parts, started
to unveil that elusive cluster population.

\begin{table*}
\caption[]{New Clusters and Associations}
\label{tab1}
\renewcommand{\tabcolsep}{8.0mm}
\renewcommand{\arraystretch}{1.25}
\begin{tabular}{lccl}
\hline\hline
Reference &Analysed Number& New Objects& Acronym \\
(1)&(2)&(3)&(4)\\
\hline
\citet{TLD88}   &  1      &        1    &     TLB\\
\citet{WB97}     &  5      &        1    &     WB\\
\citet{Walb99a}  &  2      &        2    &     WBB\\
\citet{Walb99b}   &  4      &        1    &     WDP\\
\citet{WMAB02}   &  5      &        3    &     WMB\\
\citet{HM99} &  3      &        1    &     HCD99-\\
\citet{HM00}    & 2      &        1    &     HRR\\
\citet{HM01}    &  1      &        1    &     HCD01-\\
\citet{HM01}   &  2      &        1    &     HCD02-\\
\citet{HM03}  &  2      &        2    &     HMW\\
\citet{MHMW05}  &  1     &         1    &     MHW\\
\citet{BPF05}     &  1     &         1    &     Bologna\\
\citet{Nota06}; \citet{Sabbi07}   & 16     &        16    &     NSS\\
\citet{Nakajima05}      &  25     &        18    &     NKD\\
\citet{Testor06}   &   2     &         2     &    TLF\\
\citet{Testor07}     &   2     &         2     &    TLK\\
\citet{Pietrzynski99}     & 615     &       126     &   LOGLE\\
\citet{Gouliermis03}  - Clusters         &  259     &       125     &   GKK-O\\
\citet{Gouliermis03}  - Associations      & 153     &       102    &    GKK-A\\
\citet{Gouliermis07} & 5 & 5 & GKH\\
\citet{HenneK08} & 5 & 0 & HGH\\
\citet{Schm08} &1 &1 & SGDH\\
\hline
\end{tabular}
\begin{list}{Table Notes.}
\item Otherwise stated, objects are essentially all clusters.
\end{list}
\end{table*}

\begin{figure*}
\resizebox{\hsize}{!}{\includegraphics{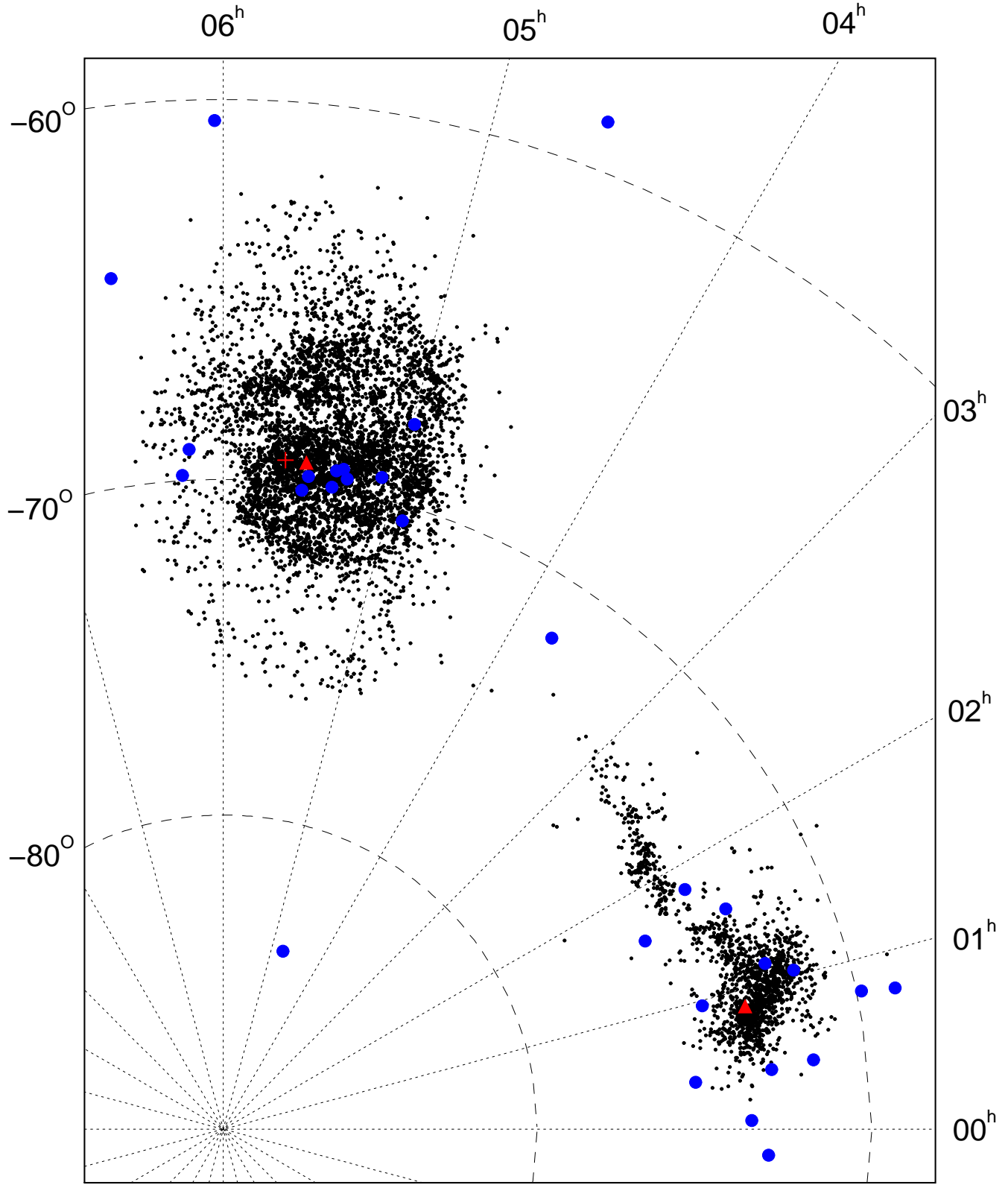}}
\caption{Angular distribution of the 9305 extended objects in the Magellanic System. Clusters older
than 4\,Gyr are shown as large blue circles. The derived LMC and SMC centroids (see Table~\ref{tabCC}) 
are indicated by red triangles. The position of 30\,Dor is shown by the plus sign.}
\label{fig1}
\end{figure*}

Bica \& Schmitt (1995, updated in \citealt{BD00}) and \citet{BSDO99} compiled general catalogues from numerous
previous catalogues and many lists sparsely distributed in the literature, and presented new findings 
based on Sky Survey plates. The total number of extended objects in these catalogues is 7900, including the 
SMC, intercloud (Bridge) region, and LMC. Cross-identifications that take into account object plate identifications, 
class, size, positions, and uncertainties were carried out.

As present-day Magellanic Cloud Surveys provide their first results, like the University of Michigan UM/CTIO 
Magellanic Cloud Emission Line Survey (MCELS)\footnote{http://www.ctio.noao.edu/mcels/ }, or are 
scheduled for soon, such as the Visible and Infrared Survey Telescope for Astronomy (VISTA)\footnote
{http://www.eso.org/gen-fac/pubs/messenger/archive/no.127-mar07/arnaboldi.pdf}, we found it timely
to update the catalogues of \citet{BSDO99} and \citet{BD00}, so that  literature objects  can be
easily identified, and new findings in deeper surveys can be confidently asserted.

Besides the update, we will use the present catalogue to investigate structural properties of the
Clouds as probed by the large-scale spatial distribution of different classes of objects. We will
also examine potential effects of the LMC, SMC, and Bridge tidal fields on the structure of individual
objects.

This paper is structured as follows. In Sect.~\ref{UpCat} we present the updates and additions to
the Magellanic System catalogue. In Sect.~\ref{GlobProp} we discuss statistical properties of the
different object classes contained in the catalogue, such as the distributions of apparent size and
ellipticity. In Sect.~\ref{DepGC} we examine dependences of the object parameters with distance to 
the Clouds centroids. In Sect.~\ref{CloudStr} we investigate the structure of both Clouds with the 
spatial distribution of the catalogue objects. Concluding remarks are given in Sect.~\ref{Conclu}.

\section{The updated Catalogue}
\label{UpCat}

The procedures used in this paper are essentially the same as those  employed in our previous 
ones. We cross-identify new and old objects by position, angular size, and object class. We give  
in Table~\ref{tab1} the statistics on clusters and associations (and related objects) from papers
published since the latest catalogue version. Most new papers deal with 
discoveries with HST. A large number of LMC clusters were studied in 
the central parts by \citet{Pietrzynski99} as part of the Optical Gravitational Lens Experiment 
(OGLE; \citealt{Udalski03}). We also included information from \citet{PU99} dealing with 
SMC clusters, and about binary and multiplet clusters in the LMC from \citet{PU00}.

\begin{figure*}
\begin{minipage}[b]{0.50\linewidth}
\includegraphics[width=\textwidth]{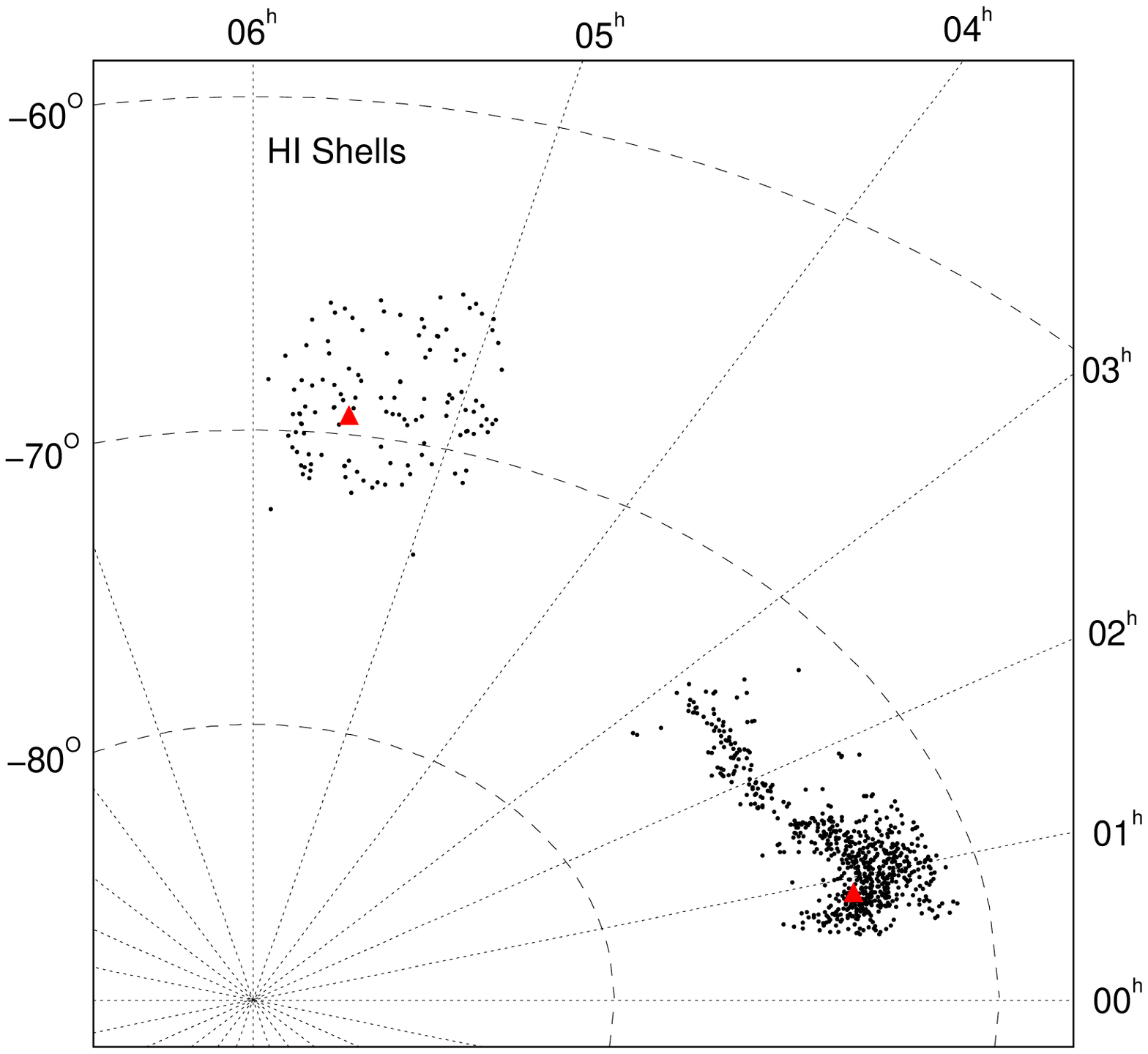}
\end{minipage}\hfill
\begin{minipage}[b]{0.50\linewidth}
\includegraphics[width=\textwidth]{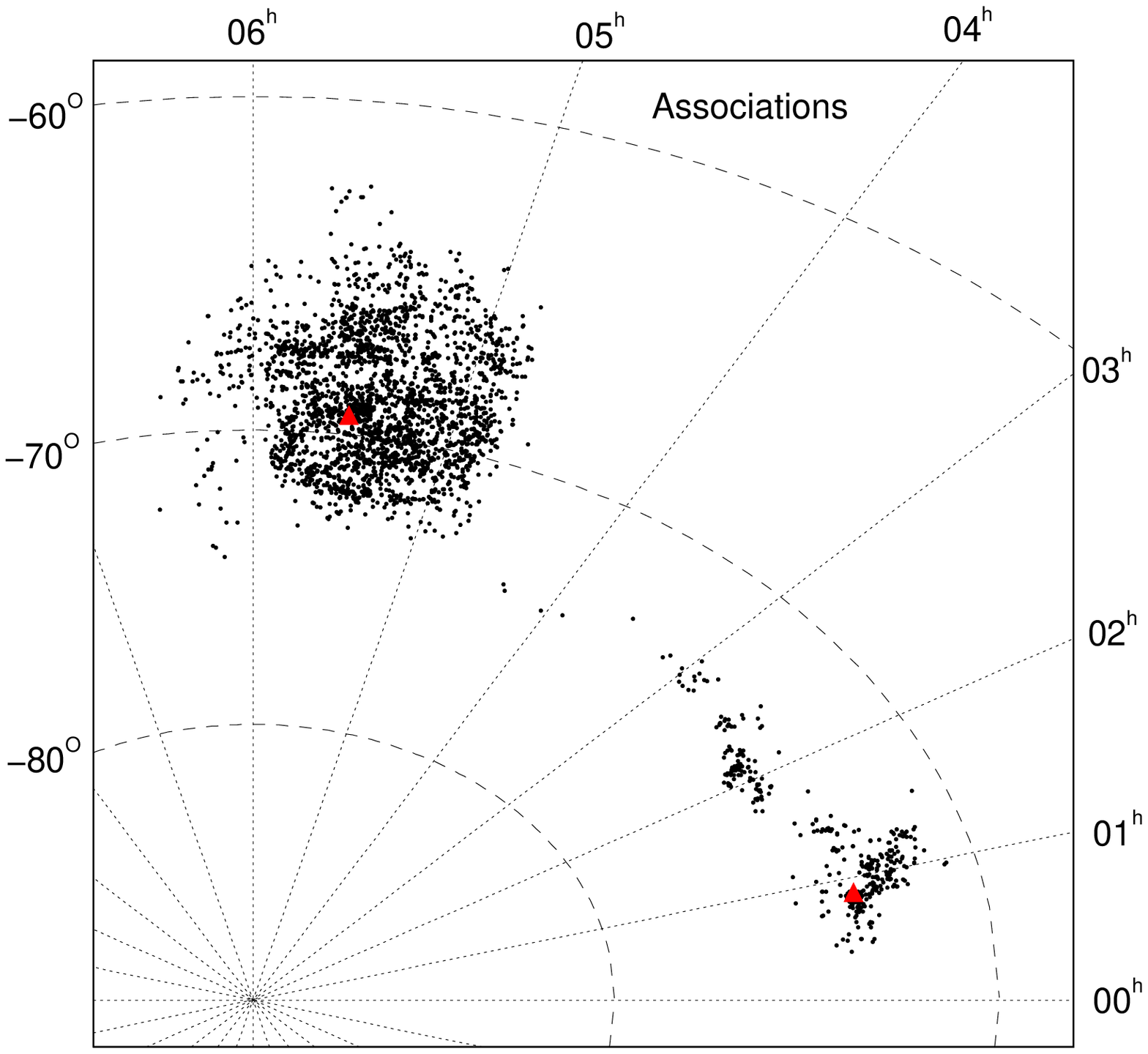}
\end{minipage}\hfill
\begin{minipage}[b]{0.50\linewidth}
\includegraphics[width=\textwidth]{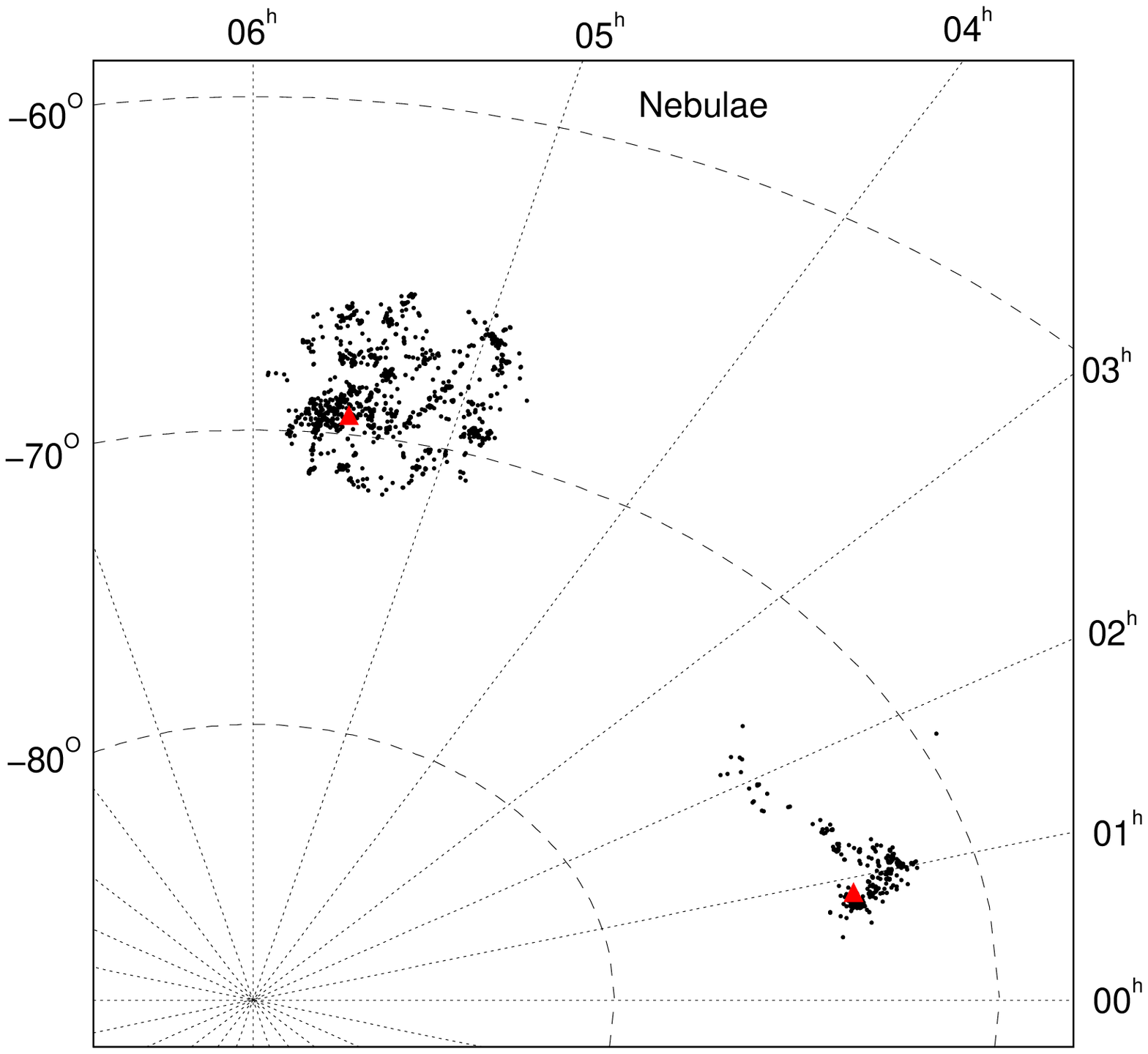}
\end{minipage}\hfill
\begin{minipage}[b]{0.50\linewidth}
\includegraphics[width=\textwidth]{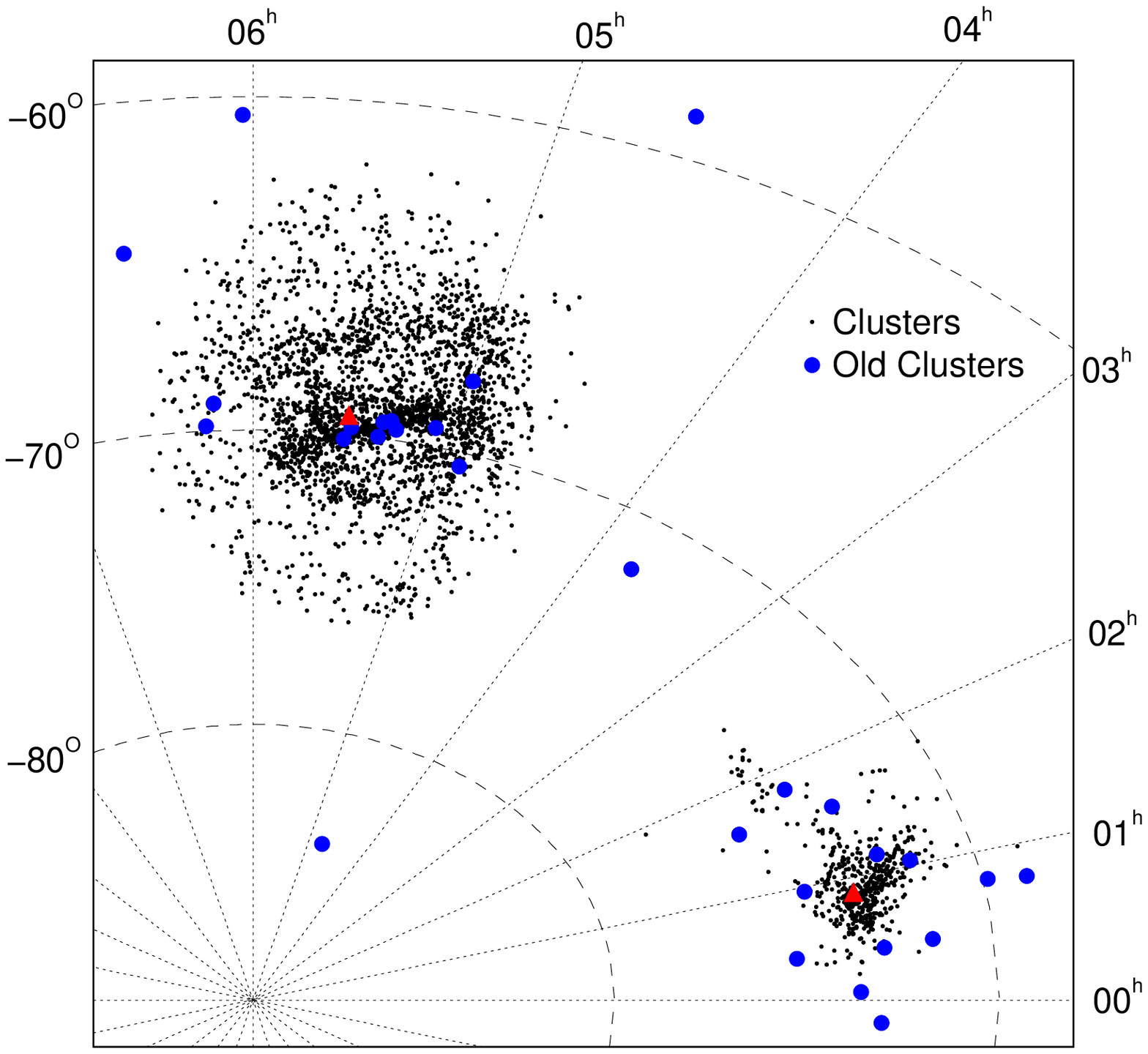}
\end{minipage}\hfill
\caption[]{Angular distribution of the HI shells and super shells (top left panel), stellar 
associations (top right), emission nebulae (bottom right), and star clusters (bottom right).
The adopted LMC and SMC centroids are indicated in all panels by filled triangles.}
\label{fig2}
\end{figure*}

Besides those, the updated catalogue includes 10 new findings (4 clusters, 3 associations,
and 3 emission nebulae) under the acronym BBDS.

A new feature of the present catalogue is the literature indications of genuine ($\rm age>9$\,Gyr) globular
clusters (GCs), with two in the SMC (\citealt{OAS87}; \citealt{MSF98}; \citealt{AAK03}) and 16 in the LMC
(\citealt{Dutra99}; \citealt{MG04}; \citealt{MPG04}). From the latter study we include ESO\,121-SC3 as an 
LMC GC, which is the only cluster in the 4-9\,Gyr LMC age gap. They suggest that ESO\,121-SC3 was accreted 
by the LMC. Also included are old SMC intermediate age clusters (IACs) at 4-9 Gyr. These studies are 
\citet{DaCosta99}, \citet{Crowl01}, \citet{Piatti01}, \citet{Piatti05}, and \citet{Piatti07}. Known GCs 
and old IACs are very useful to trace the old systems of the Magellanic Clouds (Sect.~\ref{GlobProp}).

The previous catalogue versions included 46 SNRs. Now there are 74 SNRs and candidates. The sources
were \citet{Dickel01}, \citet{LDJ03}, \citet{Heyden04}, \citet{Blair06}, \citet{WCG06}, \citet{Bojicic067},
and \citet{Chu97}.

The catalogue includes improved coordinates derived with DSS and XDSS\footnote{Extracted from the Canadian 
Astronomy Data Centre (CADC), at \em http://cadcwww.dao.nrc.ca/ } images for star clusters in the Bridge
and in the outer parts of the SMC. Several corrections were made throughout the previous 
catalogues, such as e.g. for LMC-N34A and LMC-N34B.

As another interesting case, \citet{Lindsay61} discovered the nebula L61-593 associated to an 
emission-line star in the SMC Wing. \citet{WesHen63} interpreted it as a B star with mass loss. 
With the DSS B and XDSS R images, we find that a star cluster appears to be present, now favouring 
L61-593 as an H\,II region rather than a mass-loss star.

In Table~\ref{tab2} we show the updated census of the extended objects in the Magellanic System.

In electronic form, Tables~3 to 6 contain, respectively, 3740 star 
clusters, 3326 associations, 1445 emission nebulae, and 794 HI shells and supershells. The 
SMC shells and supershells were studied by \citet{Hadz05} and \citet{Staveley97}. \citet{Hadz05}
pointed out 59 empty shells that do not appear to have stellar counterpart. Such objects are 
also indicated in Table~5. \citet{Muller03} presented shells and supershells in the Western 
part of the Bridge, while \citet{Kim99} presented those detected  in the LMC. The electronic 
tables are arranged as follows, by column: (1) - designations; (2) and (3) - the central coordinates
$\alpha(J2000)$ and $\delta(J2000)$, respectively; (4) - object class (see definitions in 
Table~\ref{tab2}); (5) and (6) - major and minor diameters ($a$ and $b$, in arcmin), respectively; 
(7) - position angle ($PA$, in degrees), with $PA=0\degr$ to the North and $PA=+90\degr$ 
to the East; and (8) - object classification, where ``mP'', ``mT'', ``m4'', and so on, mean
member of a pair, triplet, and so forth. For details see, e.g. \citet{BSDO99}. Excerpts
of the electronic tables showing the first 5 entries are given in Table~\ref{tab3_6}.

We have checked with the present catalogue spatial coincidences between clusters or
associations with the above empty shells. Except for a couple of new coincidences, the
vast majority of these shells remain empty. Interestingly, a significant fraction of the
empty shells distribute over a protuberance to the NE of the SMC, possibly an incipient
tidal tail. This protuberance shows up in the \citet{Hadz05} study. A possible interpretation 
is that these empty shells are not related to recently-formed stars. Instead, they might be 
the first stages of the gravitational collapse leading to a molecular cloud and/or to star 
cluster formation. 

Also included are H\,II regions in the Bridge that \citet{MulPar07} cross-identified
with associations from \citet{BH95}, \citet{BD00}, and probable UV ionising stellar sources
(FAUST - \citealt{Bowyer95}).

In the present study we adopt shorter acronyms for frequent objects for the sake of space, 
inspired on SIMBAD\footnote{http://simbad.u-strasbg.fr/simbad/ } designation contractions: 
SMC-DEM becomes DEMS, LMC$\_$DEM is now  DEML, 
SMC$\_$OGLE is SOGLE, and LMC$\_$OGLE  is LOGLE. In general, we adopt the authors initials as 
acronym, likewise SIMBAD. SIMBAD designations include the year of publication, and have the 
advantage to be unique, but often 
they are too long for a study like the present one. We also changed the BD designation of 
associations in the Bridge in our previous papers to ICA (intercloud association) according to 
\citet{BatDem92} and \citet{MulPar07}.

The identification of DEML\,147 as an emission nebula on the LMC bar is supported by the detection 
of a UV-bright cluster or association by \citet{Gouliermis03}, and is thus included in the present 
catalogue.

\begin{table}
\caption[]{SMC, Bridge, and LMC Extended Object Census}
\label{tab2}
\renewcommand{\tabcolsep}{0.6mm}
\renewcommand{\arraystretch}{1.25}
\begin{tabular}{lrl}
\hline\hline
Object type & Census & Comments \\
~~~~(1)&(2)&~~~~~(3)\\
\hline
Star Clusters     &    3740  &  C+CN+CA+DCN \\
C                 &    2769  &  ordinary cluster\\
CN                &      91  &  cluster in nebula\\
CA                &     861  &  cluster similar to association\\
DCN               &      18  &  decoupled cluster from nebula\\
Associations      &    3326  &  A+AN+AC+DAN\\
A                 &    1724  &  ordinary association\\
AN                &     257  &  association w/nebular traces\\
AC                &    1253  &  association similar to cluster\\
DAN               &      92  &  decoupled association from nebula\\
Emission Nebulae  &    1445  &  NA+NC+EN+SNR+DNC+DNA\\
NA                &     995  &  nebula w/embedded association\\
NC                &     260  &  nebula w/probable embedded cluster\\
EN                &       6  &  nebula wo/association/cluster\\
SNR               &      74  &  supernova remnants\\
DNC               &      18  &  decoupled nebula from cluster\\
DNA               &      92  &  decoupled cluster from nebula\\
HI shells(HS)     &     794  &  HI shells and supershells\\
\hline
\end{tabular}
\end{table}

\begin{table*}
\caption[]{Excerpts of electronic Tables~3 to 6}
\label{tab3_6}
\renewcommand{\tabcolsep}{5.0mm}
\renewcommand{\arraystretch}{1.25}
\begin{tabular}{lccccccl}
\hline\hline
Object&$\alpha(J2000)$&$\delta(J2000)$&Type&a&b&PA&Comments \\
      &(hms)&(\degr\,\arcmin\,\arcsec)&&(\arcmin)&(\arcmin)&(\degr)\\
      (1)&(2)&(3)&(4)&(5)&(6)&(7)&(8)\\
\hline
\multicolumn{8}{c}{Table~3 - Star Clusters}\\
\hline
AM-3, ESO\,28SC4 & 23:48:59 & $-$72:56:43 & C  &0.90 &0.90 & --- & Old IAC\\
L1, ESO\,28SC8   &  0:03:54 & $-$73:28:19 & C  &4.60 &4.60 & --- & Globular Cluster\\
L2               &  0:12:55 & $-$73:29:15 & C  &1.20 &1.20 & --- &\\
L3, ESO\,28SC13  &  0:18:25 & $-$74:19:07 & C  &1.00 &1.00 & --- &\\
HW1              &  0:18:27 & $-$73:23:42 & CA &0.95 &0.85 &  0  &\\
\hline
\multicolumn{8}{c}{Table~4 - Associations}\\
\hline
B3               & 0:24:00 & $-$73:38:10 &  A &1.20& 1.10 & 40\\
HW2              & 0:27:57 & $-$74:00:05 &  C &0.75& 0.55 & 70\\
H86-3            & 0:28:04 & $-$73:03:33 & AC &0.75& 0.55 & 70\\
H86-6            & 0:29:22 & $-$73:00:00 & AC &0.60& 0.45 & 20\\
HW3              & 0:29:54 & $-$73:42:03 & AC &1.50& 1.10 & 70\\
\hline
\multicolumn{8}{c}{Table~5 - Emission nebulae}\\
\hline
SMC-N3,DEMS1     & 0:31:40 & $-$73:47:43 &  NA & 1.10 & 1.10 & ---\\
DEMS2            & 0:37:15 & $-$72:59:41 & DNA & 1.80 & 1.20 & 140 & in H-A1, DC K14\\
DEMS5            & 0:41:00 & $-$73:36:22 &  NA & 2.90 & 2.90 & ---\\
DEMS6            & 0:42:14 & $-$72:59:25 &  NA & 1.10 & 1.10 & ---\\
L61-34,MA37      & 0:42:16 & $-$72:59:53 &  NC & 0.40 & 0.35 & 120 & in DEMS6\\

\hline
\multicolumn{8}{c}{Table~6 - HI shells and supershells}\\
\hline
SSH-GS1          & 0:31:26& $-$72:52:24 & HS & 5.4 & 5.4 & ---\\
SSH-GS2          & 0:32:07& $-$73:21:19 & HS & 5.8 & 5.8 & ---\\
SSH-GS3          & 0:32:15& $-$72:49:46 & HS & 2.6 & 2.6 & ---\\
SSH-GS4          & 0:33:07& $-$73:26:16 & HS &11.6 &11.6 & ---\\
SSH-GS5          & 0:33:09& $-$73:23:17 & HS & 4.8 & 4.8 & ---\\
\hline
\end{tabular}
\begin{list}{Table Notes.}
\item Col.~4: Object type as defined in Table~\ref{tab2}. Cols.~5 and 6: Major and
minor axes. Col.~7: Major axis position angle.
\end{list}
\end{table*}

With the recent additions and cross-identifications, the present catalogue contains about 
12\% more objects than those in \citet{BSDO99} and \citet{BD00} together.

Figure~\ref{fig1} shows the angular distribution of the total sample of extended objects. Outstanding
features such as the LMC central disk and bar ($PA\approx100\degr$), outer de-centred ring, the 
Bridge, and the SMC Wing and disk ($PA\approx50\degr$), have been discussed in, e.g. \citet{Wes90},
\citet{Kontizas90}, \citet{BH95}, \citet{BSDO99}, and \citet{BD00}, and references therein.

In Fig.~\ref{fig1} the old LMC clusters trace a bar-like structure, somewhat rotated with
respect to that defined by the extended objects in general. This effect was previously described
by \citet{Dot96}, where they found that the bar occupied preferentially by young clusters (SWB\,I)
is rotated with respect to the older group (SWB\,II), owing to the propagation of the perturbation
through the LMC disk that causes current star formation.

In Fig.~\ref{fig2} we show separate distributions for each individual class. We note that the LMC is
still undersampled with respect to the SMC HI shells and super-shells, and the eastern part of the Bridge 
is yet to be observed. Besides the SMC bar and Wing, the HI shells may trace additional features
possibly related to tidal effects. Interestingly, the associations suggest a spiral arm-like outer
extension in the eastern side of the LMC. The SMC disk and the Bridge are better traced by associations. 
The nebulae in the LMC appear to follow a spiral pattern centred in the 30\,Dor region
(Fig.~\ref{fig2}, lower-left panel). Finally, the old star clusters trace the LMC bar and outer 
parts, while the old SMC clusters are preferentially distributed in its outer parts.

\section{Statistical properties of the extended objects}
\label{GlobProp}

The relatively large number of objects included in the subsamples (electronic Tables~3 to 6), can be 
used to investigate  statistical properties of some
structural parameters, both in terms of object class and tidal field strength. Of
particular interest is whether effects due to the very different LMC and SMC tidal 
fields on the structural parameters and spatial distribution of the objects can be 
detected and quantified with the presently updated catalogue.

For the sake of simplicity, we separate the objects into 2 classes, {\em (i)} clusters, which contain
essentially the star clusters older than 5\,Myr, and {\em (ii)} associations and related
objects, in which we gather the HI shells and super shells, OB associations and emission 
nebulae. Besides, we also consider the spatial location of each object according 
to right ascension. We take as SMC objects those located within 
$23^h40^m<\alpha(J2000)<01^h20^m$, LMC ones at $4^h<\alpha(J2000)<6^h40^m$, while Bridge objects 
are located in between (e.g. Fig.~\ref{fig1}). We point out that the present definition of the 
Bridge is somewhat broader than that adopted in \citet{BH95}. It now includes part of the SMC 
Wing.

\subsection{Apparent diameters}
\label{Rapp}

\begin{figure}
\resizebox{\hsize}{!}{\includegraphics{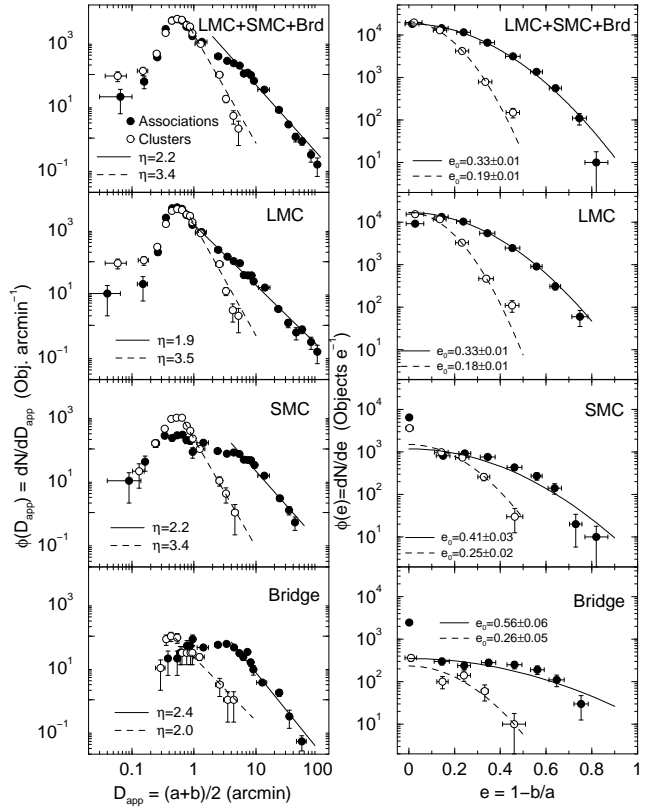}}
\caption{Left panels: apparent diameter distribution function, $\phi(\rap)=dN/d\rap$, of the
star clusters and associations. LMC, SMC, and Bridge distributions are shown separately, as well
as these three spatial structures together (top-most panel). Fits of $\phi(\rap)\sim D_{\rm app}^{-\eta}$
to the large-size tail are shown for the associations (solid line) and clusters (dashed). Right panels:
same as the left ones for the ellipticity ($e=1-b/a$) distribution function, $\phi(e)=dN/d\,e$.
Fits in the right panels correspond to the exponential-decay function $\phi(e)\sim e^{-(e/e_0)^2}$.}
\label{fig3}
\end{figure}

The updated MC catalogue gives the apparent major and minor axes, $a$ and $b$, respectively, from
which we compute the mean apparent diameter $\rap=(a+b)/2$ for each object. Based on this, we build 
the apparent diameter distribution function, defined as $\phi(\rap)=dN/d\rap$. In Fig.~\ref{fig3} 
(left panels) we show $\phi(\rap)$ of the clusters and associations located in the LMC, SMC, Bridge, 
as well as in the MC system as a whole. Star clusters and associations present different distributions
in all MC subsystems. In particular, associations tend to have objects with larger diameters
than the clusters.


Another interesting fact is that the apparent diameter distributions fall off as a power-law
for objects that occupy the large-size tail (Fig.~\ref{fig3}). Indeed, the distributions can
be reasonably well fitted with the function $\phi(\rap)\sim\rap^{-\eta}$ in the range
$D_{\rm app}^{min}\leq\rap\leq D_{\rm app}^{max}$ (Fig.~\ref{fig3}). Table~\ref{tabRap}
summarises the fit details. The number of LMC and SMC star clusters fall off towards large
diameters at a faster rate ($\sim D_{\rm app}^{-3.4}$) than the associations ($\sim D_{\rm app}^{-2}$),
while in the Bridge the slopes are similar ($\sim D_{\rm app}^{-2}$). The slope in the apparent
radii distribution of the associations agrees with that predicted (and measured) for H\,II regions
in spiral galaxies (\citealt{Oey03}). Since most of the clusters are significantly older than
the associations, the difference in slope (and maximum size) probably reflects the several Myr
of dynamical evolution and disruption effects operating on the former structures. Besides, the
steeper decline with apparent diameter observed in the LMC and SMC cluster $\phi(\rap)$, with respect
to the Bridge, is consistent with the stronger tidal field of the Clouds.

If extrapolated to the small-radii tail, the decaying-power law distribution of apparent
diameters in the LMC, SMC, and Bridge (top-left panel in Fig.~\ref{fig3}), would suggest that 
the number of observed objects represents a small
fraction of the total population. Indeed, because of the difference in slope, the fraction of observed
associations would be $\sim2.7\%$, and $\sim0.6\%$ for the clusters. Known small clusters are in
general embedded in HII region complexes (Table 1). Their small number certainly stresses the fact
that systematic surveys for small-scale structures are yet to be carried out.

\setcounter{table}{6}
\begin{table}
\caption[]{Properties of the large-size tail of $\phi(\rap)$}
\label{tabRap}
\renewcommand{\tabcolsep}{2.25mm}
\renewcommand{\arraystretch}{1.25}
\begin{tabular}{lcccc}
\hline\hline
Reference&$D_{\rm app}^{min}$&$D_{\rm app}^{max}$ &$\phi_0$&$\eta$ \\
Sample&(\arcmin)&(\arcmin) \\
\hline
All assoc.   &4.4&100&$7119\pm1583$&$2.2\pm0.1$ \\
All clusters &0.76&5.3&$1774\pm160$&$3.4\pm0.2$\\
\hline
LMC assoc.   &0.95&100&$1439\pm87$&$1.9\pm0.1$ \\
LMC clusters &0.76&5.3&$1516\pm162$&$3.5\pm0.2$  \\
\hline
SMC assoc.   &5.4&43&$3181\pm1098$&$2.2\pm0.1$ \\
SMC clusters &0.76&4.6&$209\pm7$&$3.4\pm0.1$ \\
\hline
Bridge assoc.   &4.4&55&$1789\pm687$&$2.4\pm0.2$ \\
Bridge clusters &0.54&4.4&$22\pm3$&$2.0\pm0.2$ \\
\hline
\end{tabular}
\begin{list}{Table Notes.}
\item Fits with the function $\phi(\rap)=\phi_0\,D_{\rm app}^{-\eta}$ are performed for
$D_{\rm app}^{min}\leq\rap\leq D_{\rm app}^{max}$. The combined LMC, SMC, and Bridge samples
are represented by the `All' reference sample.
\end{list}
\end{table}

\begin{figure}
\resizebox{\hsize}{!}{\includegraphics{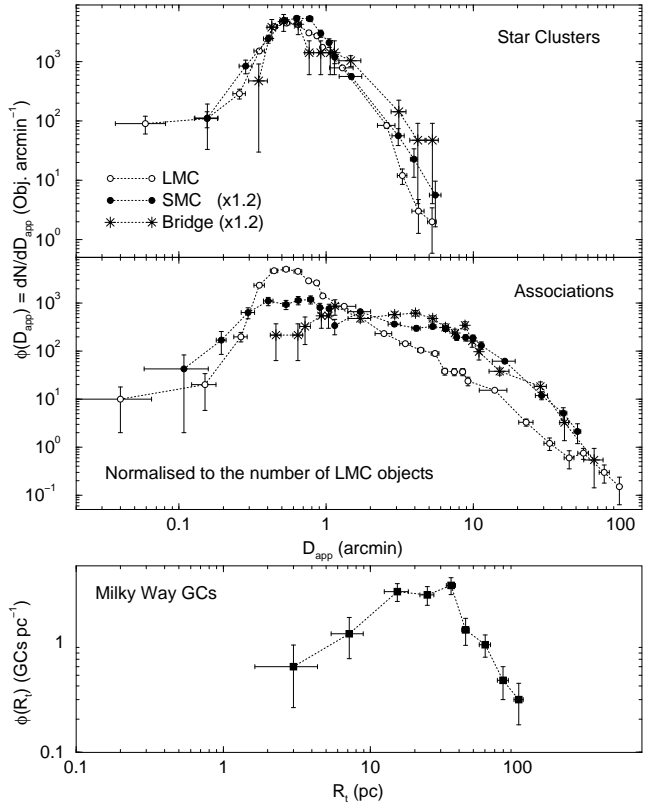}}
\caption{Top panels: apparent diameter distribution function of the LMC, SMC, and Bridge built
with star clusters (top) and associations (middle). SMC and Bridge apparent diameters have 
been multiplied by 1.2 to correct for the different distances with respect to the LMC. These 
functions have been normalised to the LMC number of objects. Bottom panel: tidal radius distribution 
function of the Milky Way GCs, in absolute scale. For a consistent comparison with the MCs apparent 
diameters, the dynamical range of the abscissa is equal in all panels. }
\label{fig4}
\end{figure}

The apparent diameter distribution functions (normalised to the same number of objects for 
inter-class comparisons) of similar classes of objects in the LMC, SMC, and Bridge, are shown in
Fig.~\ref{fig4} (top panels). For a more intrinsic analysis, SMC and Bridge apparent diameters
have been multiplied by 1.2, to account for the different distances with respect to the LMC 
(Sect.~\ref{intro}). Within uncertainties, the star clusters present similar distributions,
especially in the LMC and SMC. With the available data, the Bridge appears not to harbour clusters
smaller than $\rap\la0.3\arcmin$. As for the HI shells, associations and nebulae, the SMC and Bridge
present similar distributions, and both appear to have an excess of objects larger than $\rap\approx2\arcmin$
with respect to the LMC. This effect may be associated to the weaker SMC and Bridge tidal fields, which
allow the presence of distended, low-binding energy objects, such as those included in the association
class. The LMC and SMC distributions present a steep drop towards smaller \rap, beginning at
$\rap\approx0\farcm5$. At the LMC and SMC distances, this corresponds to physical radii of $\approx4$\,pc.
Such clusters (or associations) are not small by Galactic open cluster standards. In fact, this corresponds
to average-size Galactic open clusters (see, e.g. Fig.~7 in \citealt{FSR1767}). This raises the question of
whether such a drop is a real effect associated to formation processes and/or dissolution, an observational
limitation linked to completeness, or more probably, a combination of both. In any case, the completeness
is not the same in the 3 MC subsystems. Because of the lower surface brightness of the background and
the less-populous nature of the Bridge, star clusters and associations stand out more, and completeness 
effects in the Bridge are expected to be less important than in the Clouds.

At this point, it may be interesting to compare the MCs apparent diameter distribution functions
with the equivalent one built with the Galactic population of GCs, which is basically complete
and probes all the old Galactic substructures (see, e.g. \citealt{FSR1767}). Obviously, the Galactic
GCs are essentially old systems, while the MCs distribution functions contain young objects as
well. However, the main purpose here is to examine the shape of the Milky Way (MW) GC size-distribution
function, especially at the small-size tail. In principle, it should be more correct to include the
Galactic open clusters in this analysis, since they span a wide range in ages and populate especially
the young tail of the age distribution. However, contrary to the GCs, the open clusters are severely
affected by completeness, especially at the faint-end of the luminosity distribution (e.g. \citealt{DiskProp}),
which might introduce a completeness-related drop towards small open clusters in the size distribution
function.

Thus, with the above arguments in mind, we take as reference of GC size the tidal radii
given by Harris (1996, and the update in 2003\footnote{\em http://physun.physics.mcmaster.ca/Globular.html}).
Additionally, we consider as well the tidal radii of 11 faint GCs (not included in \citealt{Harris96})
derived by \citet{Struc11GCs}, and the recently studied GC FSR\,1767 (\citealt{FSR1767}). Since
MCs objects are essentially at the same distance from the Sun, the MW GC tidal radii are
converted to the parsec scale for a consistent comparison. The latter conversion is based on the
updated GC distances to the Sun given by \citet{GCProp}. For comparison purposes, the dynamical
range of the MW GCs tidal radii have matched to the angular scales of the MCs (Fig.~\ref{fig4}).

The tidal radii distribution function of the MW GCs is shown in Fig.~\ref{fig4} (bottom panel).
Qualitatively, it presents similar features as those of the MCs objects, especially the
relatively narrow width of the MC star clusters distribution function. Besides a maximum
between $16\la\rt(\rm pc)\la30$, the distribution function of the Galactic GCs drops off
both towards small and large radii. If the MW GCs sample is indeed basically complete, this suggests
that the small-size drop observed in the MCs distribution functions may be real, at least in part.

The peak distribution of apparent diameters in the MC system occurs for $\rap=0\farcm53 - 0\farcm77$ 
which, for an average distance of $\approx55$\,kpc, corresponds to radii in the range $\approx4.2 - 6.2$\,pc. 
Such radii are a factor $\sim4 - 5$ smaller than the peak tidal radii of the MW GCs. Most of the
difference may be accounted for by the fact that we deal with apparent sizes (measured on images as far
as the background limit) in the MC system and tidal radii (which comes from, e.g. a King-profile fit) in 
the MW. Although most of the cluster stars can be considered to be contained 
inside the apparent radius, it is smaller than the tidal radius. For instance, the tidal radii 
computed for populous and relatively high Galactic latitude MW OCs such as M\,67, NGC\,188, and 
NGC\,2477, are about 4 times larger than the respective apparent radii (\citealt{DetAnalOCs}).

\subsection{Ellipticity}
\label{ellip}

We apply a similar analysis to the ellipticity ($e=1-b/a$) distribution function,
$\phi(e)=dN/d\,e$. LMC, SMC, and Bridge clusters follow similar distribution functions
(Fig.~\ref{fig3}, right panels), especially the LMC and SMC ones. The fractional number 
of clusters decreases monotonically with ellipticity within the range $0.03\la e\la0.45$,
in all structures considered.

While the ellipticity distribution function of the clusters is similar in the LMC, SMC, and
Bridge, the associations, on the other hand, have different properties in different spatial 
structures. In particular, LMC and SMC associations contain objects with higher ellipticity 
values ($0.03\la e\la0.8$) than those of the clusters, with a slower decay of the fractional 
number with increasing $e$ (Fig.~\ref{fig3}). 

Bridge associations have an ellipticity distribution that is rather flat in the
range $e\sim0.1-0.5$, which indicates the presence of an important fraction of
non-circular objects. Besides, the distribution reaches a peak for the nearly circular objects.
SMC associations follow a similar, although less flat, distribution. LMC associations, on the other
hand, follow a smoothly-decreasing distribution, with a peak at $e\approx0.15$, and dropping
somewhat at $e=0$. Besides, LMC, SMC, and Bridge associations reach significantly higher values of
$e$ than the corresponding clusters, $e\la0.8$.

Analytically, the combined LMC$ + $SMC$ + $Bridge ellipticity distribution functions for the 
associations and clusters (Fig.~\ref{fig3}, top-right panel) are well described by the function 
$\phi(e)\sim e^{-(e/e_0)^2}$, with dimensionless ellipticity scales $e_0=0.33\pm0.01,~0.19\pm0.01$, 
respectively. Given the relative similarity of the remaining, isolated distribution functions in 
the LMC, SMC, and Bridge with the combined ones, it is obvious that they follow the same analytical 
function, but with different ellipticity scales. Indeed, the best-fit functions (Fig.~\ref{fig3}, 
right panels) are obtained for association-ellipticity scales about twice larger than those of the 
clusters. While the cluster ellipticity scale increases $\approx44\%$ from the LMC to the SMC and 
Bridge, $e_0$ for the associations increases by $\approx70\%$.

The above aspects are consistent with the fact that associations, in general, tend to be
systems less-bound than clusters, and thus more subject to distortions by tidal fields.
Besides, strong tidal fields may prevent the survival to advanced ages of a large
population of distended objects, because of induced torques, tidal disruption, and so on. 
With time, such effects may either disrupt significantly-distended objects, or at least, 
make them more circular at later ages. In any case, the qualitative aspects of the ellipticity 
distribution functions, as well as the environmental dependence of the ellipticity-scale, 
appear to correlate with the relative strengths of the LMC, SMC, and Bridge tidal fields.

\subsection{Position angle and alignment}
\label{PA}

\begin{figure}
\resizebox{\hsize}{!}{\includegraphics{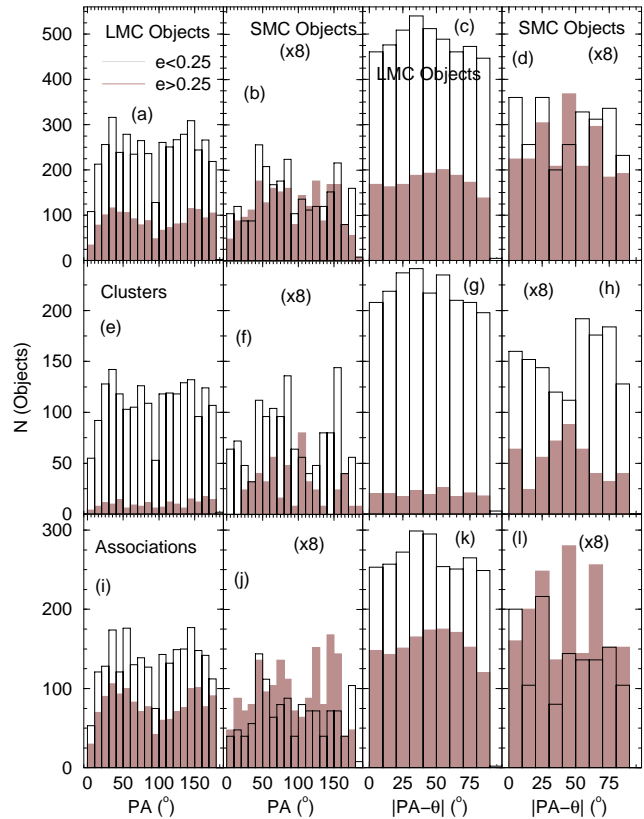}}
\caption{Histograms of the Position Angle and Alignment Angle measured in the LMC and SMC objects.
Ellipticities larger (shaded histograms) and smaller (white) than $e=0.25$ are considered
separately. Note that the SMC histograms have been multiplied by 8 for visualisation
purposes.}
\label{fig5}
\end{figure}

In Fig.~\ref{fig5} we examine the distribution of the position angle ($PA$) of the clusters
and associations around both Clouds. Objects with ellipticity higher and lower than $e=0.25$
are considered separately. In the LMC, $\approx92\%$ of the clusters with $PA$ measured are
more circular than $e=0.25$ (panel e), while in the SMC this fraction drops to $\approx74\%$
(f). As for the associations, the corresponding fractions are $\approx63\%$ in the LMC (i) and
$\approx42\%$ in the SMC (j). Clusters, in both Clouds, tend to be more circular than the 
associations, especially in the LMC, which again is consistent with the respective tidal field
strengths of the Clouds, and the relative binding energy of the objects. Interestingly, there 
is a significant drop in the number of LMC objects with $PA\approx100\degr$, especially the 
associations, but conspicuous as well for the $e\la0.25$ clusters. Since the $PA$ of the LMC bar
(Fig.~\ref{fig1}) is on average $\approx100\degr$ (Fig.~\ref{fig1}), one might speculate whether
there is an enhanced dissolution of objects with $PA$ parallel to the bar. We expect that the maximum
tidal effect will occur for extended objects at the tips of the bar with parallel $PA$, which might
imply a resonant effect. The collapse of molecular clouds is expected to be essentially radial,
so that clusters do not acquire much rotation during formation. Indeed, the bar/$PA$ alignment does 
not occur in the SMC, probably because of its less prominent bar.

We also estimate the alignment between each object's $PA$ and the angle defined by direction vector
($\theta$) with respect to the corresponding Cloud centroid, $|PA-\theta|$. The angle $\theta$ is
measured in the same way as $PA$ (Sect.~\ref{UpCat}). Since the alignment is symmetrical with respect 
to the sign of $PA-\theta$, and corresponds to the smaller angle, the measured values are in the range
$0\degr\leq|PA-\theta|\leq90\degr$. Clusters and associations of both Clouds (Fig.~\ref{fig5}, 
panels c, d, g, h, k, and l) do not appear to present statistically significant trends in 
$|PA-\theta|$.

\section{Parameters as a function of distance to the centroids}
\label{DepGC}

Most Milky Way GCs have a size that scales with the Galactocentric distance (e.g. \citealt{vdBMP91};
\citealt{Struc11GCs}). Part of this relation may have been established as early
as at the Galaxy formation, when the higher density of molecular gas in central regions may have
produced smaller clusters (e.g. \citealt{vdBMP91}). Dynamical evolution, especially that driven by
external processes such as tidal disruption, collision with giant molecular clouds, disk and spiral
arms, is important as well to establish a relation of increasing cluster size with Galactocentric
distance. Such processes lead to the disruption of most star clusters in a mass-dependent time-scale
shorter than $\approx1$\,Gyr (\citealt{Gieles06}). Since the latter effects are more critical for
low-mass objects located close to strong tidal fields, a similar relation has been observed for the
Galactic open clusters (e.g. \citealt{Lynga82}; \citealt{Tad2002}; \citealt{DetAnalOCs}; \citealt{OldOCs}).

\begin{figure}
\resizebox{\hsize}{!}{\includegraphics{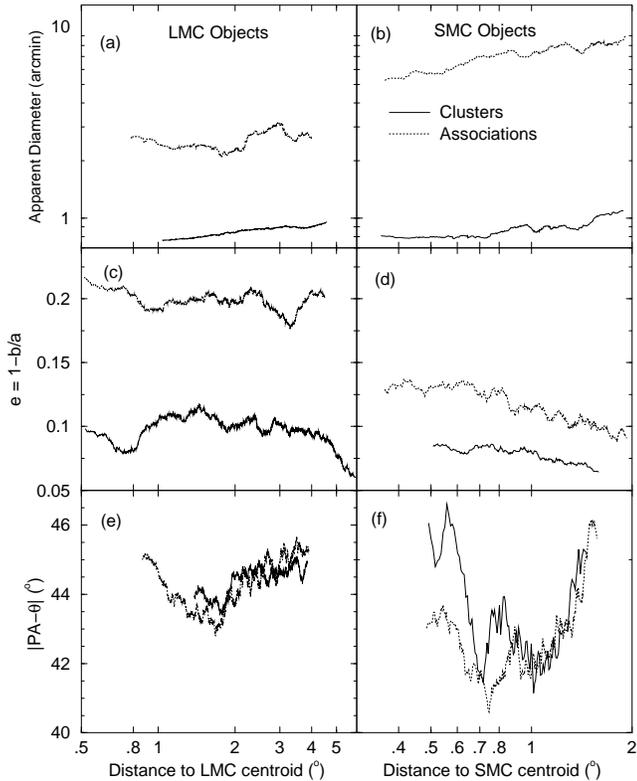}}
\caption{The average apparent diameter (top panels), ellipticity (middle) and alignment angle 
(bottom) of the LMC (left panels) and SMC (right) objects are examined as a function of the 
distance to the respective Cloud centroid. Curves correspond to fiducial lines. SMC apparent 
diameters have been converted to the values corresponding to the LMC distance. }
\label{fig6}
\end{figure}

In the top panels of Fig.~\ref{fig6} we investigate the above issue with the apparent diameters of the
catalogue LMC and SMC objects (converted to the values corresponding to the LMC distance). The main
purpose here is to search for trends, thus to minimise scatter we work with running averages, which 
correspond to the average value of a given parameter within bins usually containing 20\% of the number 
of objects in each sample. The result of this procedure are the fiducial lines shown in Fig.~\ref{fig6}. 
While LMC and SMC clusters have, on average, similar sizes, SMC associations are larger, as already implied 
by Fig.~\ref{fig4}. Clusters and associations in both Clouds appear to follow a trend of increasing apparent 
size with angular distance to the respective centroid, except perhaps for some fluctuation in the LMC 
associations.

The ellipticity of the SMC objects presents a similar relation with distance to the centroids, in the
sense that objects closer to each Cloud's centroid tend to be less circular (middle-right panels 
of Fig.~\ref{fig6}). However, this relation is milder in the LMC objects.

Finally, there appears to exist a mild correlation between the alignment angle and distance
from the centroid for the LMC star clusters, in the sense that clusters closer to the LMC have
the $PA$ more aligned with the direction axis (bottom panels of Fig.~\ref{fig6}). The alignment, in 
this case, occurs at the angle $|PA-\theta|\approx45\degr$. A similar relation appears for the 
LMC associations and SMC associations and clusters, but only for objects more distant than 
$\approx1.5\degr$ (LMC) and $\approx0.7\degr$ (SMC). The trend appears to be reversed for
objects closer than these distances. However, as a caveat we note that such trends are rather 
speculative, since variations of $\approx3\degr$ and $\approx6\degr$ in $|PA-\theta|$ of the
LMC and SMC, respectively, may be within the measurement uncertainties. 

\section{Large-scale structure of the MC system}
\label{CloudStr}

We analyse the large-scale structure of the Clouds by means of the angular distribution of extended
objects (adding star clusters and associations in general). Since the Clouds do not have symmetrical
structures (Fig.~\ref{fig1}), we employ two different approaches in what follows. First we explore
angular slices along nearly perpendicular directions (Sect.~\ref{Azim}), and concentric radial 
distribution (Sect.~\ref{RDP}).

\subsection{Azimuthal extractions}
\label{Azim}

\begin{figure}
\resizebox{\hsize}{!}{\includegraphics{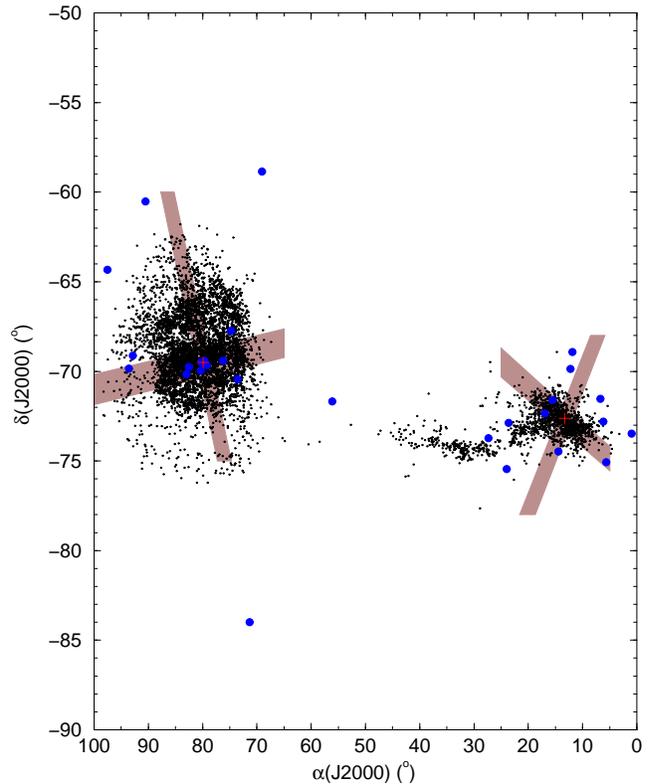}}
\caption{Geometry of the extracted angular slices. The wider stripe in the LMC traces the bar
and beyond, while in the SMC it follows the disk/bar direction. Nearly-perpendicular slices 
were also extracted. Slices are centred on the geometrical centroids (plus signs) of the LMC 
bar and SMC bar/disk structure. Clusters older than 4\,Gyr are shown as blue circles.}
\label{fig7}
\end{figure}

The position of the extracted slices take advantage of the direction of the prominent LMC bar, and the 
possible bar/disk structure. The adopted geometry of the extractions is illustrated in Fig.~\ref{fig7}. 
We examine the structure along the LMC bar and in a perpendicular direction, which probes the underlying
disk. In the SMC we extract slices along the disk/bar and nearly perpendicular directions. As reference, 
we take as centres of the slices the geometrical centroids of the LMC bar and the SMC disk/bar 
(Table~\ref{tabCC}).

\begin{figure}
\resizebox{\hsize}{!}{\includegraphics{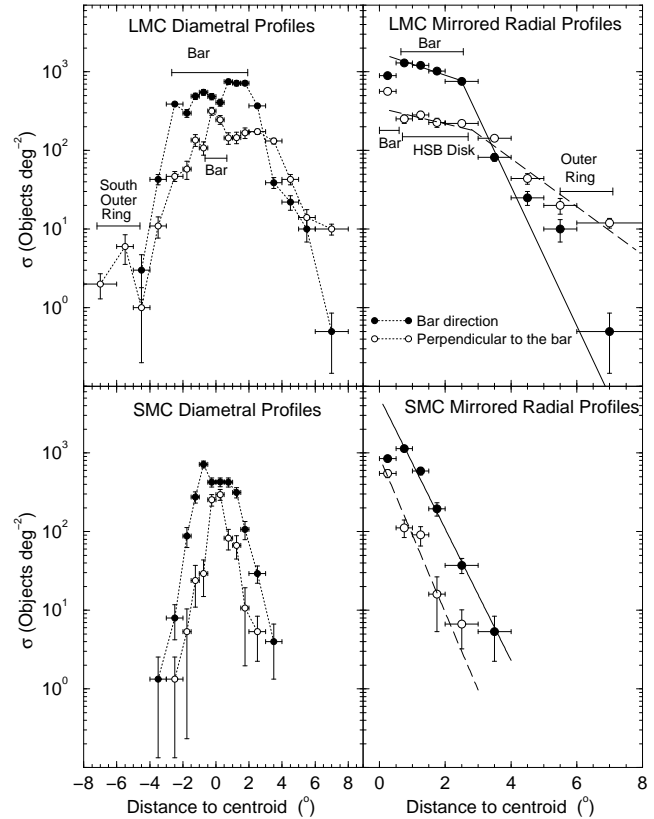}}
\caption{Left panels: azimuthal profiles extracted along the LMC bar and SMC bar/disk directions
(filled circles). Profiles extracted in nearly-perpendicular directions are also shown (empty circles).
Right panels: radial profiles built from the azimuthal ones by folding them over the centroid. Main
LMC structures are indicated. Fits with exponential disks are shown.}
\label{fig8}
\end{figure}

Figure~\ref{fig8} shows the azimuthal density profiles for the LMC (top-left panel) and
SMC (bottom-left). The distance of a given object with the observed equatorial coordinates 
$(\alpha,\delta)$ to the adopted centroid $(\alpha_0,\delta_0)$, is computed as
$R=\sqrt{((\alpha-\alpha_0)\cos(\delta))^2 + (\delta-\delta_0)^2}$. Thus, because of the
declination $\delta\approx-70\degr$ of the Clouds, angular separations in Fig.~\ref{fig8}  
correspond to $\approx1/3$ of those implied by the axes of Fig.~\ref{fig1}.

As expected, the main LMC structures, such as the bar, the high-surface
brightness (HSB) disk, and the outer ring, can be detected in the profiles. The SMC profiles
appear to be described essentially by exponential disks. While both SMC
profiles are almost symmetrical with respect to the adopted centroid, the prominent LMC
structures introduce significant asymmetries, both with respect to the bar centroid and between
the perpendicular slices. The LMC bar is a factor $\approx5$ denser (in terms
of the number of objects per area) than the HSB disk. The central excess in the perpendicular
LMC profile corresponds to part of the bar. In the SMC, the average density of the bar/disk
profile is a factor $\approx5$ denser than that of the nearly-perpendicular direction.

In any case, we build as well pseudo-symmetrical profiles by folding the opposite sides of the 
azimuthal extractions over the respective centres. In this process, symmetrical points that 
occur at the same bin of distance to centroid are averaged out. The mirrored radial profiles are 
shown in the right panels of Fig.~\ref{fig8}, where the main structural features can be seen as well. 
We fit these profiles with an exponential-disk function adapted to number counts,
$\sigma(R)=\sigma_{0D}\times e^{-(R/R_D)}$, where $\sigma_{0D}$ represents the number-density of
objects at the centre, and $R_D$ is the disk scale length. As expected, both SMC profiles are well
fit by exponential-disks, with the scale lengths $R_D=0\degr.52\pm0\degr.03\rm\approx0.54\pm0.03\,kpc$
and $0\degr.44\pm0\degr.08\rm\approx0.46\pm0.08\,kpc$, respectively for the bar/disk parallel and
nearly-perpendicular profiles. The LMC profiles, on the other hand, require two exponential-disks each
to account for the dense structures. Thus, the profile parallel to the bar can be represented by the
disk scale lengths $R_D=3\degr.1\pm0\degr.3\rm\approx2.7\pm0.3\,kpc$ and
$~0\degr.49\pm0\degr.05\rm\approx0.43\pm0.04\,kpc$, for the bar region and beyond the bar
limits, respectively. In the nearly-perpendicular direction, two disks are also necessary, but in this
case the inner one (within the bar limits) corresponds to the HSB disk. In this case, the number-density
excesses over the disk profile correspond to the additional outer disk (Fig.~\ref{fig1}). We derive
$R_D=4\degr.4\pm1\degr.2\rm\approx3.8\pm1.0\,kpc$ and $1\degr.42\pm0\degr.15\rm\approx1.2\pm0.1\,kpc$
for the inner and outer parts of the perpendicular profile. The innermost points (Fig.~\ref{fig8}) in both
profiles were excluded from the fits.

There is evidence that the Magellanic Clouds have complex spatial structure. Indeed, the LMC has 
an inclined disk-like structure (e.g. \citealt{Olsen02}), which is warped (e.g. \citealt{Nikolaev04}), 
and flared (e.g. \citealt{Alves00}). Besides, the deprojected LMC structure is elliptical, being more
extended along the North-South direction (\citealt{vdM01}). The SMC appears to have a 3-dimensional 
structure more extended along the line-of-sight (\citealt{Crowl01}). The present analysis (Fig.~\ref{fig8}) 
is based on the angular distribution of the objects in both Clouds. Thus, it is possible that part of 
the differences detected in the two perpendicular directions in both Clouds arises from their intrinsic
properties and projection effects.

\subsection{Concentric radial distribution}
\label{RDP}

Alternatively, we analyse the spatial distribution of the extended objects by means of the 
radial density profiles (RDPs). The RDPs corrrespond to the projected radial surface-density, i.e, 
the number-density of objects contained in concentric rings around the LMC and SMC centroids. 
The underlying assumption for this kind of analysis, which is mostly applied to star clusters,
is that the structures should present an important degree of radial symmetry. This is not the 
case of the Clouds, as discussed in previous sections. In any case, RDPs still can be used as 
probes of the radial distribution of objects averaged over all azimuthal directions and, consequently, 
of the large-scale structure.

Table~\ref{tabCC} gives the LMC and SMC centroid coordinates derived for the spatial distribution
of the combined clusters$+$associations (and related objects), as well as those corresponding to the
clusters separately. These centroids correspond to the region where the maximum number-density of objects
in each class occurs. In the LMC the cluster centroid is shifted $\approx7\arcmin$ and $\approx3\arcmin$
to the West and South, respectively, with respect to the combined distribution centroid. The offset in
the SMC is about twice as large, but in the opposite directions. We recall that the position of 30\,Dor
(R\,136) is $\alpha(J2000)=05^h38^m42^s$, and $\delta(J2000)=-69\degr06\arcmin02\arcsec$, thus somewhat
to the East and North of the centroids in Table~\ref{tabCC} (Fig.~\ref{fig1}).

\begin{table}
\caption[]{Centroid coordinates of the different reference systems}
\label{tabCC}
\renewcommand{\tabcolsep}{2.3mm}
\renewcommand{\arraystretch}{1.25}
\begin{tabular}{lcccc}
\hline\hline
Reference&$\alpha(J2000)$&$\delta(J2000)$&$\Delta\alpha$&$\Delta\delta$\\
System&(hms)&($\degr\,\arcmin\,\arcsec$)&(\arcmin)&(\arcmin)\\
\hline
LMC combined&05:31:28.8&$-$69:22:30&---&---\\
LMC clusters&05:31:00.0&$-$69:22:48&$-7.2$&$-3.0$ \\
\hline
SMC combined&00:52:31.2&$-$73:15:00&---&---\\
SMC clusters&00:53:28.8&$-$73:07:48&$+15.0$&$+7.2$\\
\hline
LMC bar&05:19:36.6&$-$69:07:08&$-$178&$-$7.7\\
SMC bar/disk&00:53:10.3&$-$72:37:12&$+9.8$&$+37.8$\\
\hline
\end{tabular}
\begin{list}{Table Notes.}
\item The offsets of the centroid coordinates derived from the cluster distribution
with respect to the combined (clusters$+$associations) distribution are given by 
$\Delta\alpha$ and $\Delta\delta$. The last two lines correspond to the geometrical 
centroids of the LMC bar and SMC bar/disk.
\end{list}
\end{table}

It is worth noting that a centroid definition depends on which tracer is used. For instance,
\citet{deV72} obtained $\alpha(J2000)=5^h23^m24^s$ and $\delta(J2000)=-69\degr44\arcmin00\arcsec$ as 
the optical centre of the LMC bar. With stellar density contours in the infrared, \citet{vdM01} 
obtained $\alpha=5^h25^m05^s$ and $\delta=-69\degr47\arcmin00\arcsec$ as the LMC centre. Finally,
\citet{Kim98} found the LMC HI kinematic centre to lie at $\alpha=5^h17^m24^s$ and 
$\delta=-69\degr02\arcmin00\arcsec$. Thus, our LMC bar centroid (Table~\ref{tabCC}), which is 
particularly sensitive to the distribution of young clusters and associations, lies somewhat to the 
North-West of the optical value provided by \citet{deV72}. The present LMC centroid (Table~\ref{tabCC}) 
lies to the East of those of \citet{vdM01} ($\approx6^m$) and \citet{Kim98} ($\approx14^m$), and 
halfway ($\approx22\arcmin$) between them in declination. As for the SMC, \citet{Wes90} reported
$\alpha(J2000)=0^h49^m47^s$ and $\delta(J2000)=-72\degr55\arcmin40\arcsec$ as the optical centre,
which lies $\approx15\arcmin$ and $\approx3^m$ to the North-West of the centroid derived in this
work (Table~\ref{tabCC}). 

\begin{figure}
\resizebox{\hsize}{!}{\includegraphics{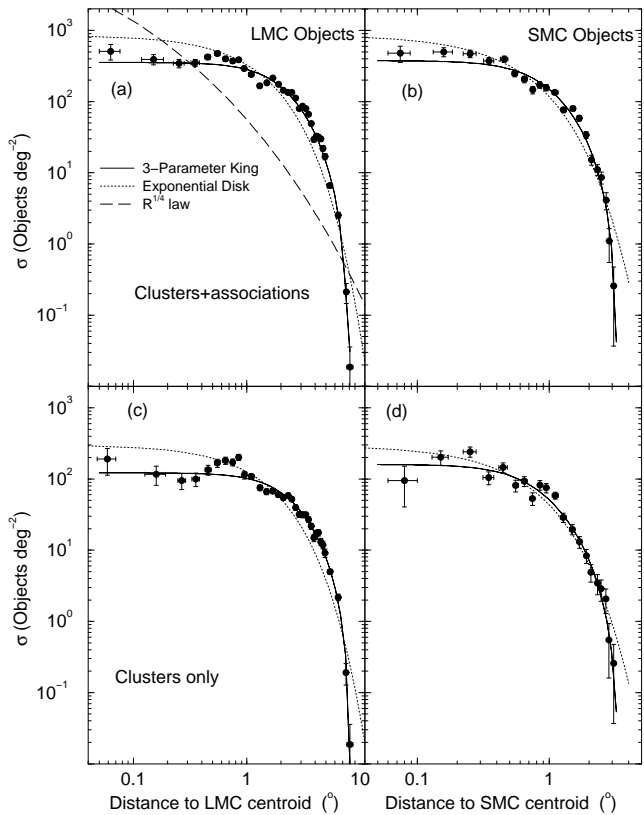}}
\caption{Radial density profiles for the clusters and associations combined (top panels)
and clusters separately (bottom), located in the LMC (left) and SMC (right). Fits with the
3-parameter King (solid line) and exponential disk (dotted) profiles are shown in all
panels. Panel (a) contains a tentative fit with the $R^{1/4}$ law (dashed line).}
\label{fig9}
\end{figure}

We build the RDPs with the centroid coordinates derived for the combined distributions (Table~\ref{tabCC}).
They are shown in Fig.~\ref{fig9}, with the combined clusters$+$associations (top panels) 
and clusters, separately (bottom). Many of the catalogue objects are young, and the 
corresponding RDPs should represent the relatively recent spatial distribution, $t\la200$\,Myr 
(\citealt{Bica96}).

At first sight, the RDPs (Fig.~ \ref{fig9}) present similar shapes, with a relatively flat and 
extended central region followed by a steep decline towards large galactocentric radii. As expected 
from its larger size, the LMC RDPs reach a distance of $R\approx8\degr$ from the centre, while in
the SMC, $R\la3\degr$.

As a first approach to describe the LMC and SMC structures implied by the RDPs, by means of an 
analytical function, we test an exponential-disk profile. In all cases, the overall fit fails to 
reproduce important RDP features. In particular, it overestimates the density of objects especially 
in the central parts and external region, and underestimates it in the mid region. Fit parameters 
are given in Table~\ref{tab4}. In any case, this kind of fit suggests that the LMC disk-scale length 
is $\approx1\degr$, about twice the SMC value. We also try the $R^{1/4}$ law (\citealt{deV48}), but 
as shown for the LMC RDP (panel a), it fails completely. Obviously, a combination of both would not 
either describe the profiles. One conclusion, drawn from such a statistically comprehensive catalogue, 
is that the (angular-average) large-scale structure of both interacting irregular galaxies does not 
follow the classical disk and/or spheroidal laws.

Alternatively, we also test a 3-parameter profile based on the \citet{King66} law, which usually 
describes the structure of star clusters, especially the Galactic GCs and open clusters, by means 
of the surface-brightness distribution. Formally, we express the adopted King-like profile as
$\sigma(R)=\sigma_{0K}\left[1/\sqrt{1+(R/R_c)^2}-1/\sqrt{1+(R_t/R_c)^2}\right]^2$, where
$R_c$ and $R_t$ are the core and tidal radii, respectively, and $\sigma_{0K}$ is the central
density of objects. Qualitatively, the corresponding profiles reproduce well the observed RDPs
over the full radial range (Fig.~\ref{fig9}). Within uncertainties, the structural parameters
implied by the combined clusters$+$associations and clusters alone RDPs are similar (Table~\ref{tab4}).
For the LMC we derive a core radius $R_c\approx2.6\degr$ which, at the LMC distance ($\approx50$\,kpc),
corresponds to $R_c\approx2.3$\,kpc; for the tidal radius we derive $R_t\approx8.1\degr\approx7$\,kpc.
Similar considerations for the SMC ($\approx60$\,kpc) lead to $R_c\approx1\degr\approx1$\,kpc,
and $R_t\approx3.3\degr\approx3.5$\,kpc. Thus, in absolute units, the SMC structural radii correspond 
to about half of the LMC ones. Finally, the concentration parameters $c_p=\log(\rt/\rc)$, in 
all cases, are comparable to those of the least concentrated Galactic GCs (see, e.g. Fig.~7
in \citealt{FSR1767}).

King profiles usually describe the structure of virialised systems, such as the old Galactic GCs,
but many of the LMC and SMC objects used in the above analysis are young. However,
we note that the structure of some young Galactic star clusters have been shown to follow the King
profile as well, e.g. the $\sim1.3$\,Myr open cluster NGC\,6611 (\citealt{N6611}), and NGC\,4755
(\citealt{N4755}), with $\sim10$\,Myr of age. Ironically, the best-fit to the average radial
distribution of objects in both Clouds is given by the King-like profile, which is not usually
applied to the structure of galaxies.

\begin{table*}
\caption[]{Structural parameters measured in the RDPs with the 3-parameter King and exponential-disk
profiles (Fig.~\ref{fig9})}
\label{tab4}
\renewcommand{\tabcolsep}{3.3mm}
\renewcommand{\arraystretch}{1.25}
\begin{tabular}{lccccccccc}
\hline\hline
&\multicolumn{5}{c}{$\sigma_{0K}\left[1/\sqrt{1+(R/R_c)^2}-1/\sqrt{1+(R_t/R_c)^2}\right]^2$}
&&\multicolumn{3}{c}{$\sigma_{0D}\times e^{-(R/R_D)}$}\\
\cline{2-6}\cline{8-10}
RDP&$\sigma_{0K}$&\rc&\rt&$c_p$&CC&&$\sigma_{0D}$&$R_D$&CC\\
   &$\rm(Obj.\,deg^{-2})$&(deg)&(deg)& & &&$\rm(Obj.\,deg^{-2})$&(deg)&\\
(1)&(2)&(3)&(4)&(5)&(6)&&(7)&(8)&(9)\\
\hline
LMC combined&$705\pm30$ &$2.4\pm0.2$&$7.9\pm0.2$&0.52&0.982&&$860\pm108$&$1.0\pm0.05$&0.898\\
LMC clusters&$261\pm11$ &$2.8\pm0.2$&$8.4\pm0.2$&0.48&0.984&&$303\pm39$&$1.1\pm0.05$&0.897 \\
SMC combined&$820\pm62$ &$1.1\pm0.1$&$3.3\pm0.1$&0.48&0.978&&$861\pm115$&$0.5\pm0.03$&0.978 \\
SMC clusters&$293\pm27$ &$0.9\pm0.1$&$3.4\pm0.2$&0.58&0.967&&$295\pm37$&$0.5\pm0.03$&0.947 \\
\hline
Old clusters&$1.4\pm0.3$&$4.0\pm1.5$&$11.0\pm0.5$&0.44&0.912&&$1.2\pm0.3$&$1.9\pm0.2$&0.943 \\
\hline
\end{tabular}
\begin{list}{Table Notes.}
\item Col.~5: concentration parameter, $c_p=\log(\rt/\rc)$. Col.~6: fit correlation
coefficient. Combined: star clusters$+$associations.
\end{list}
\end{table*}

\begin{figure}
\resizebox{\hsize}{!}{\includegraphics{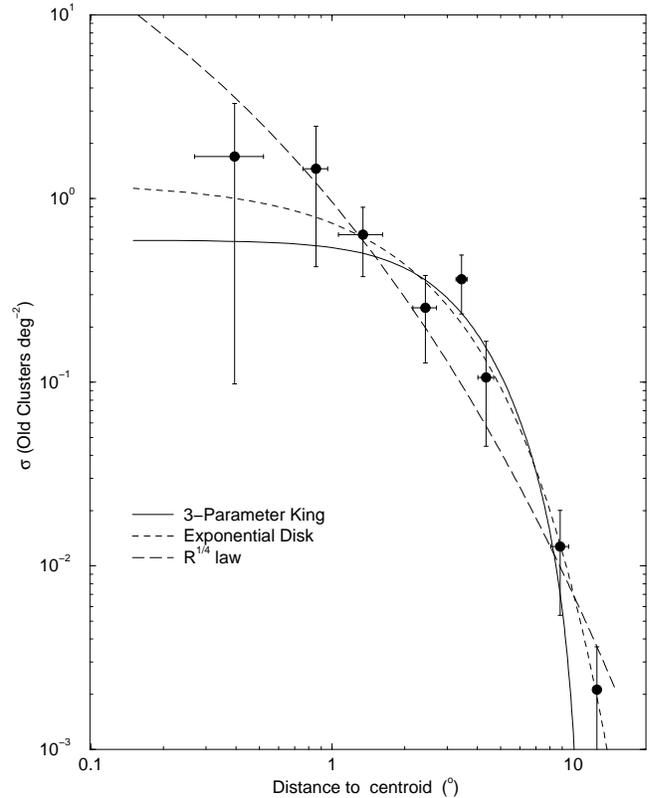}}
\caption{Same as Fig.~\ref{fig9} for the old star clusters of the LMC and SMC. The
LMC and SMC individual RDPs have been merged to increase the statistics. The exponential 
disk appears to be the best fit to the RDP.}
\label{fig10}
\end{figure}

Each Cloud harbours about half of the 29 old star clusters present in the catalogue (electronic
Table~3; Fig.~\ref{fig1}). Such clusters are much older than the time elapsed since the last encounter
between both Clouds (e.g. \citealt{BekChi07}), and they can be used to probe whether the
present-day spatial distribution of the old clusters retains information on the early-Cloud
structure. Two of these are the very distant GCs NGC\,1841 and Reticulum, at
$R\approx10\degr$ (Fig.~\ref{fig1}), which are likely LMC members (\citealt{Suntz92}). We consider
for this analysis a composite RDP, in which we compute for each old star cluster the distance
to the nearest MC centroid. The combined LMC and SMC RDP is shown in Fig.~\ref{fig10}. At first sight, it
appears to represent a more extended structure than those discussed in Fig.~\ref{fig9}, but
the extension may be caused by the two distant GCs. In any case, the exponential disk, with
a scale-length of about $2\degr$, now appears to describe the full radial range of the RDP
somewhat better than the 3-parameter King profile, while the $R^{1/4}$ law fails altogether.
The more extended character of this RDP is reflected on the fit structural parameters, which
are larger than those derived with the younger objects (Table~\ref{tab4}).

\section{Summary and conclusions}
\label{Conclu}

The primary goal of this paper is to update the catalogue of extended objects in the
Magellanic System. With the recent addition of HST, CCD mosaics, and survey data, the 
number of known objects in the Clouds now reaches 9503, which represents a relatively 
substantial increase of $\approx12\%$ with respect to the previous versions 
(\citealt{BH95}; \citealt{BSDO99}; \citealt{BD00}). It now includes HI shells and supershells,
cross-identifications with the previous literature, and subsequently discovered objects.

Such a number of objects is large enough to allow for a statistically significant analysis of
environmental effects on the distribution of structural parameters among the different classes
of objects, in the LMC, SMC, and Bridge tidal fields separately. Star clusters present similar
distributions of structural parameters in the three MC subsystems. SMC associations (and related
objects, emission nebulae, and HI shells and supershells), on the other hand, tend to be larger
and more circular than in the LMC. We also detect evidence that the apparent diameter of clusters
and associations increase with the distance to each Cloud centroid. The ellipticity presents the
opposite trend, especially in the SMC. These relations are consistent with the relative strengths
of the LMC, SMC, and Bridge tidal fields. Indeed, the standard model of \citet{BekChi07} assumes the
masses $M_{\rm LMC}=2\times10^{10}\,\ms$ and $M_{\rm SMC}=3\times10^{9}\,\ms$ for the LMC and SMC,
respectively. Obviously, the Bridge is less massive than the SMC. Such masses can produce significant
tidal stress on star clusters and associations, depending on their location.

With respect to the angular distribution of objects, number-density profiles extracted along
the LMC bar and in a perpendicular direction can be reasonably well represented by two exponential
disks, one for the bar region and the other for the outer parts of the profile (Fig.~\ref{fig8}). 
Disk scale-lengths parallel to the bar are $\approx2.7$\,kpc and $\approx0.4$\,kpc, for the bar 
region and beyond, respectively. In the perpendicular direction they are $\approx3.8$\,kpc and 
$\approx1.2$\,kpc, for the high-surface brightness disk and outer ring, respectively. Similar 
profiles extracted along equivalent directions in the SMC follow a single exponential disk with 
$\approx0.5$\,kpc of scale-length. 

Alternatively, when (angular-averaged) radial number-density profiles are considered, the
large-scale structure of both Clouds appears to be best described by a 3-parameter King-like
function, characterised by core and halo sub-structures. In this case, the LMC core and tidal 
radii are $\rc\approx2.6\degr$ and $\rt\approx8.2\degr$, respectively; SMC values are a factor
$\approx0.4$ of the LMC ones. In absolute scale, LMC values are $\rc\approx2.3$\,kpc and 
$\rt\approx7.2$\,kpc, while SMC ones are about half of these. The tidal/core radii ratio in 
both Clouds imply low concentration parameters, comparable to those of sparse 
Galactic GCs.

What emerges from the present work is a scenario where the present-day, (angular-averaged) 
large-scale structures of both Clouds appear to behave as tidally-truncated systems (which is 
not unexpected, since they are Milky Way satellites), characterised by well-defined core and 
halo sub-structures. This picture comes about despite the fact that both Clouds are not spherical 
systems (Fig.~\ref{fig1}). Thus, they have undergone severe tidal perturbation when the last dynamical 
and hydrodynamical interaction between the Clouds took place, about 200\,Myr ago (\citealt{BekChi07}). 
Taken isolately, the older LMC and SMC star clusters, on the other hand, appear to be distributed 
as an exponential disk. This distribution is possibly reminiscent of the Clouds structure prior to 
the last interaction.

The LMC/SMC interaction is not unusual in the local Universe. They can be classified into the
minor-merger interaction picture, which involves high and low-mass galaxies (e.g. \citealt{FerPas04},
and references therein). A series of examples of such interactions involving galaxies with prominent
bulge and disk has been studied in that paper. Some of the disk galaxies developed the double-disk
structure, similarly to the present case of the LMC.

\section*{Acknowledgements}
We acknowledge partial support from CNPq (Brazil). We thank the anonymous referee for interesting
suggestions.


\end{document}